\documentclass[final,english]{bullsrsl}[2022/06/15]

\usepackage[latin1]{inputenc}
\usepackage[T1]{fontenc}

\usepackage{natbib} 
\usepackage{graphicx}

\begin{document}
\title{Characterization of Eclipsing Binary System EPIC 211982753}

\author[affil={1,2}, corresponding]{Alaxender}{Panchal}
\author[affil={1}]{Yogesh}{Chandra Joshi}
\affiliation[1]{Aryabhatta Research Institute of observational sciencES (ARIES), Nainital, 263001, India}
\affiliation[2]{Department of Physics, DDU Gorakhpur University, Gorakhpur, 273009, India}
\correspondance{alaxender@aries.res.in}
\date{19th July 2023}
\maketitle


%

\begin{abstract}
We present the physical parameters of an eclipsing binary system EPIC 211982753 derived through photometric and radial velocity data modeling. We make use of photometric data from NASA's K2 mission, ASAS-SN, and 1.3-m Devasthal Fast Optical Telescope (DFOT) while spectroscopic data have been acquired from the HERMES spectrograph at 1.2-m Mercator telescope. The linear ephemeris for the system is updated using the K2 mission data. The synthetic lightcurve and radial velocity curves are generated with the help of eclipsing binary modeling package PHOEBE 1.0. The masses of primary and secondary components are determined as 1.64 $\pm$0.02 and 1.55 $\pm$0.01 $M_{\odot}$, respectively. The radius for primary and secondary components are estimated as 1.73 $\pm$0.02 and 1.47 $\pm$0.02 $R_{\odot}$, respectively. The distance of the system is calculated as 238 $\pm$ 4 pc. The eclipsing binary is found to be a total eclipsing system with a high mass ratio of q=0.94. 

\end{abstract}


\section{Introduction}
Eclipsing binaries (EBs) hold an important place in astronomy as they provide a simple but accurate way to measure stellar parameters as compared to single stars. The periodic changes in their flux and radial velocity (RV) are used to estimate the physical parameters for these systems. The applications of double-lined detached EBs for understanding star formation and evolution are discussed by many authors \citep{2010A&ARv..18...67T, 2017A&A...608A..62H, 2021A&A...647A..90F}. The physical parameters of the components determined using EBs are more precise as compared to other techniques such as gravity-mass or isochrones fitting. The EBs can also be used to determine distances with uncertainties up to 3$\%$ \citep{2005A&A...429..645S, 2020ApJ...904...13G}. Using long-term photometric observations, the orbital period variation can be studied which is further used to understand/detect the component interaction (mainly in close systems), magnetic activity cycles, or extra component in the system \citep{1958BAN....14..131K, 2021AJ....161..221P, 2021AJ....162...13L, 2022ApJ...927...12P}.

Due to multiple space missions like Kepler \citep{2010Sci...327..977B} and TESS \citep{2015JATIS...1a4003R}, a large amount of precise photometric data is available for thousands of stars. The Kepler Eclipsing Binary Catalog (Third Revision) and TESS Eclipsing Binary Catalog consist of photometric time series data for $\sim$2800 and $\sim$4500 systems, respectively. In this work, we present the physical parameters of an EB system EPIC 211982753 which was first reported by \cite{2011MNRAS.416.2477W} after analyzing NASA's STEREO mission data. The system was again observed by the K2 mission and \cite{2016AA...594A.100B} mentioned it as an EB candidate on the basis of K2 campaign-5 (C-5) observations.

\section{Observations}
Photometric time series data are available for the target in the Kepler archive (can be accessed through the Barbara A. Mikulski Archive for Space Telescopes) and ASAS-SN. We also observed the system in $R_{c}$ band using DFOT, Nainital \citep{2022JAI....1140004J}.
The available K2 LCs are already corrected for instrumental systematics and spacecraft pointing errors using EPIC Variability Extraction and Removal for Exoplanet Science Targets (EVEREST) pipeline. EVEREST pipeline uses pixel-level decorrelations and Gaussian processes for this purpose.
Furthermore, this source was also observed in surveys like WISE, GAIA (G, GBP, and GRP bands) but with poor coverage or bad data quality. To detect the radial velocity (RV) variation, we collected 10 high-resolution spectra for the system using the High-Efficiency and high-Resolution Mercator Echelle Spectrograph (HERMES) at the Mercator Telescope (La Palma, Spain). HERMES spectra were reduced using the HERMES data pipeline. Three medium-resolution spectra (MRS) were collected from $6^{th}$ data release (DR6-v2) of the Large Sky Area Multi-Object Fibre Spectroscopic Telescope (LAMOST).  

\section{Updating Ephemeris}
Photometric observations by the K2 mission are used to update the ephemeris of the system. The system was observed under C-5 (April 28, 2015 to July 10, 2015) and C-18 (May 14, 2018 to July 2, 2018). The times of minimum brightness (ToMs) at primary and secondary eclipses are determined via parabola fitting. The O-C from ToMs are calculated at different orbital cycles (E) using a period of 5.389920 days and BJD 2457141.728936 as ToM at $0^{th}$ orbital cycle (BJD$_{\circ}$). For the O-C diagram, we determined 17 primary ToMs and 18 secondary ToMs using K2 data. \cite{2011MNRAS.416.2477W} and \cite{2018MNRAS.477.3145J} also reported primary ToM as 2454274.329535 and 2456381.80891 but we did not use these ToMs in the present analysis due to poor data quality and unknown uncertainty. The updated linear ephemeris and associated uncertainties are determined using MCMC implemented in EMCEE \citep{2013PASP..125..306F}. The updated linear ephemeris is estimated as follows:



\begin{equation}
\label{li_epic1}
BJD_{\circ} (E)=2457141.7291(\pm 0.0004)
+5.389917(\pm 0.000003)\times E
\end{equation} 

\begin{figure}
\begin{center}
\includegraphics[width=8cm,height=8cm]{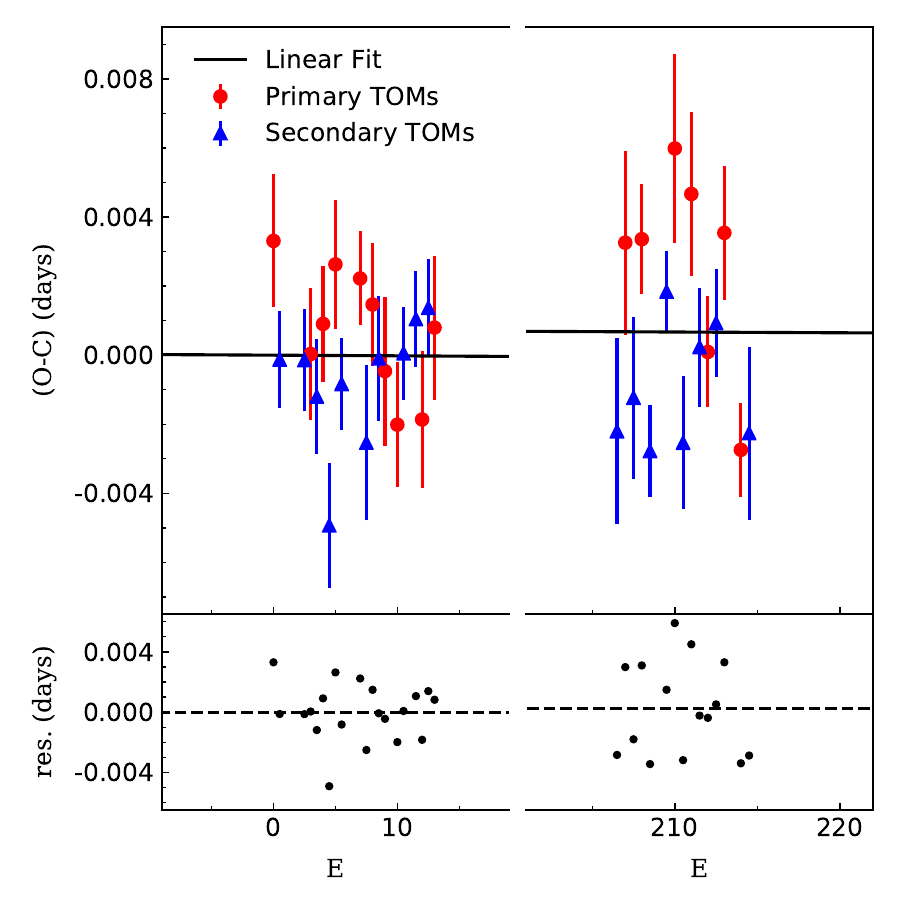}
\includegraphics[width=7.5cm,height=8cm]{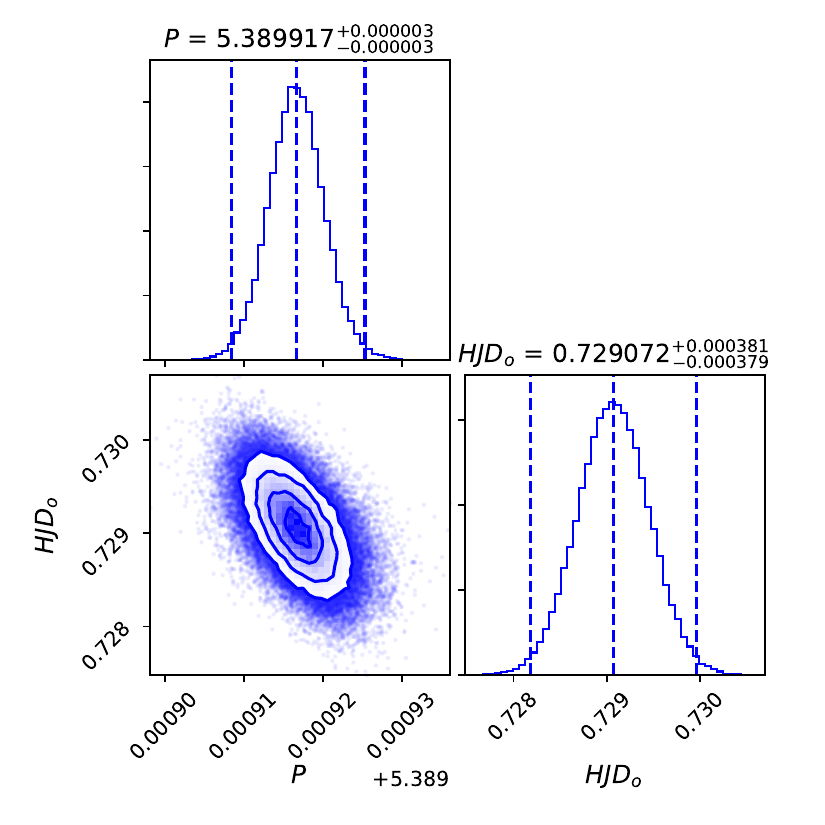}
\caption{The O-C diagram with linear fit (left panel) and the corner plot for updated ephemeris (right panel). The $HJD_{o}$ in the right panel of the Figure is $HJD_{o}$-2457141.}
\label{fig_01}
\end{center}
\end{figure}

The O-C diagram with a linear fit is shown in the left panel of Figure~\ref{fig_01}. The O-C diagram region from E=20 to E=200 is excluded due to the unavailability of data points. The fitted line shows a discontinuity in the O-C diagram due to the small slope of the fitted line and excluded region. The O-C diagram follows a linear trend which can be represented by:
\begin{equation}
\label{li_epic_oc1}
(O-C) =-0.237836(\pm 508.62) \times10^{-6}
-1.89206(\pm 3.68891) \times10^{-6}\times E
\end{equation}

Red dots and blue triangles represent the O-C for primary and secondary ToMs. The right panel of  Figure~\ref{fig_01} represents the MCMC distribution for the updated ephemeris.
The quadratic fit for the O-C diagram is as follows:


\begin{equation}\label{li_epic_oc2}
(O-C) =0.00003(\pm 0.00094)
      -0.000007(\pm 0.000121)   \times E
      +0.23(\pm 5.54) \times10^{-7}\times E^2
\end{equation}

The O-C diagram fit does not show any significant change in the $\chi^{2}$ due to quadratic fit. Because of significantly large errors on the quadratic term in the (O-C) fit, the period of the system can be considered constant. 

%
\section{Modeling and Parameter Determination}
\begin{figure}
\begin{center}
\includegraphics[width=7.5cm,height=7cm]{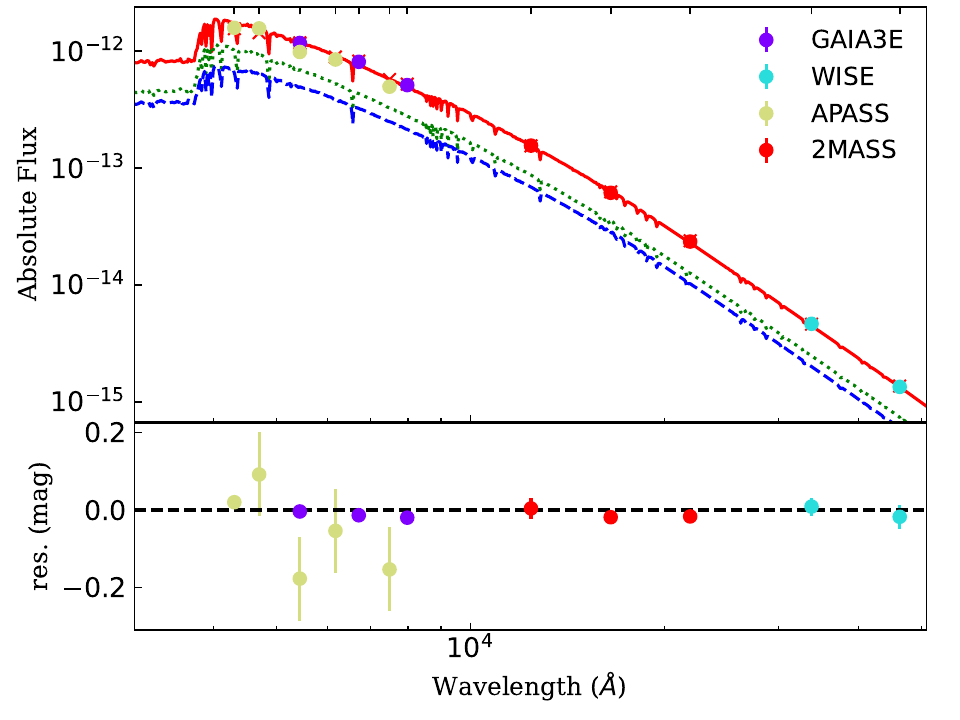}
\includegraphics[width=7.5cm,height=7cm]{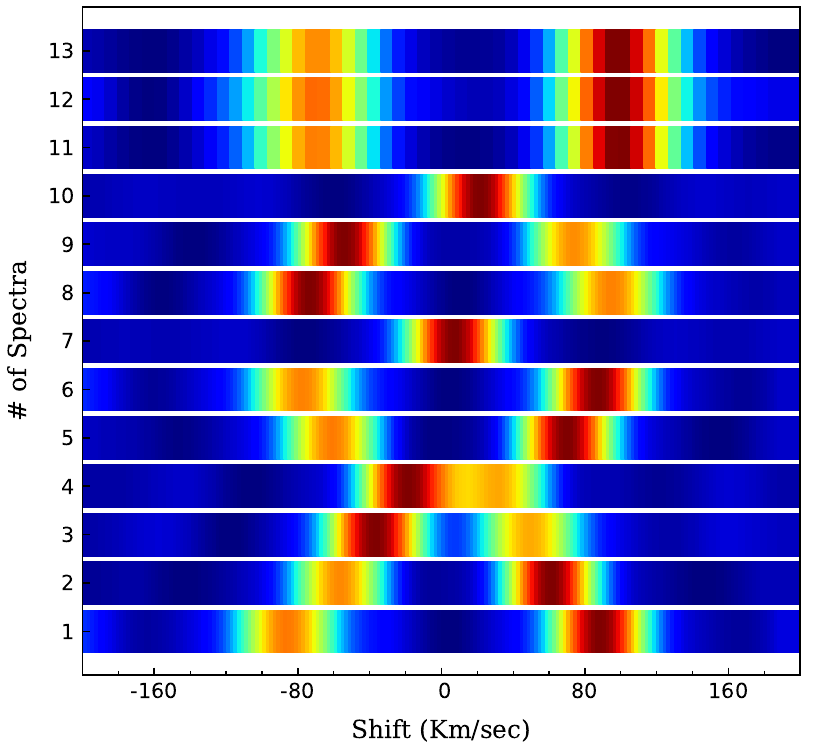}
\caption{The fitted SED with SPEEDYFIT is shown in the left panel. The flux contributions from individual components are shown by dotted and dashed lines. The continuous line represents the combined SED of the system. The right panel shows the CCFs for different spectra in the form of a heatmap.}
\label{fig_02}
\end{center}
\end{figure}

\textbf{SED Fitting: } Spectral energy distribution (SED) provides an ideal method to independently determine the stellar parameters. For this purpose, a freely available SED fitting tool SPEEDYFIT was used. The published photometry information was used to generate the SED for the system. For GAIA and WISE, we calculated and used out-of-eclipse flux of the system in the SED. The distance was obtained through the parallax given by GAIA DR3 while reddening was taken from \citet{2011ApJ...737..103S}. The SPEEDYFIT uses \cite{1979ApJS...40....1K} model grids and MCMC approach to determine the SED parameters along with the uncertainties. The primary component SED (green dotted line) and secondary component SED (blue dashed line) are shown in Figure~\ref{fig_02}. The photometric measurements in different photometric bands are shown with filled circles in different colors. From the SED fitting, the primary and secondary component $T_{eff}$ are determined as 7243$^{+103}_{-160}$ and 7122$^{+114}_{-138}$ K.

%

%
\textbf{Radial Velocity Curve: } To determine the radial velocity (RV) from the observed spectra (HERMES and LAMOST), the FXCOR routine of IRAF was used. FXCOR utilizes the Fourier cross-correlation technique to compare the target and template spectra. A synthetic template spectrum was generated using the SPECTRUM software package \citep{1999ascl.soft10002G}. Stellar atmospheric models by \cite{2003IAUS..210P.A20C} were used during synthetic spectra generation. The projected rotational velocity was fixed at 25 km/sec as reported by \cite{2020AJ....160..120J} using APOGEE-2 spectra. The $\log g$ and [Fe/H] for the synthetic spectra were kept at 4.0 and 0.0, respectively. The temperature for the synthetic template was taken as 7250 K (close to the SED temperature estimates). The right panel of Figure~\ref{fig_02} shows the cross-correlation functions (CCFs) for all the spectra in the form of a heatmap. The color variation from blue to red describes the height of CCF. The CCF numbers 1-10 are generated from HERMES spectra while 11-13 are generated from LAMOST MRS spectra. The primary peak in CCF is visible in dark red while the secondary peak falls around orange region. The RVs were calculated by fitting double Gaussian to the actual CCFs. For observations close to primary/secondary eclipse, only a single peak is observable in CCFs.

\begin{figure}
\begin{center}
\includegraphics[width=15cm,height=8cm]{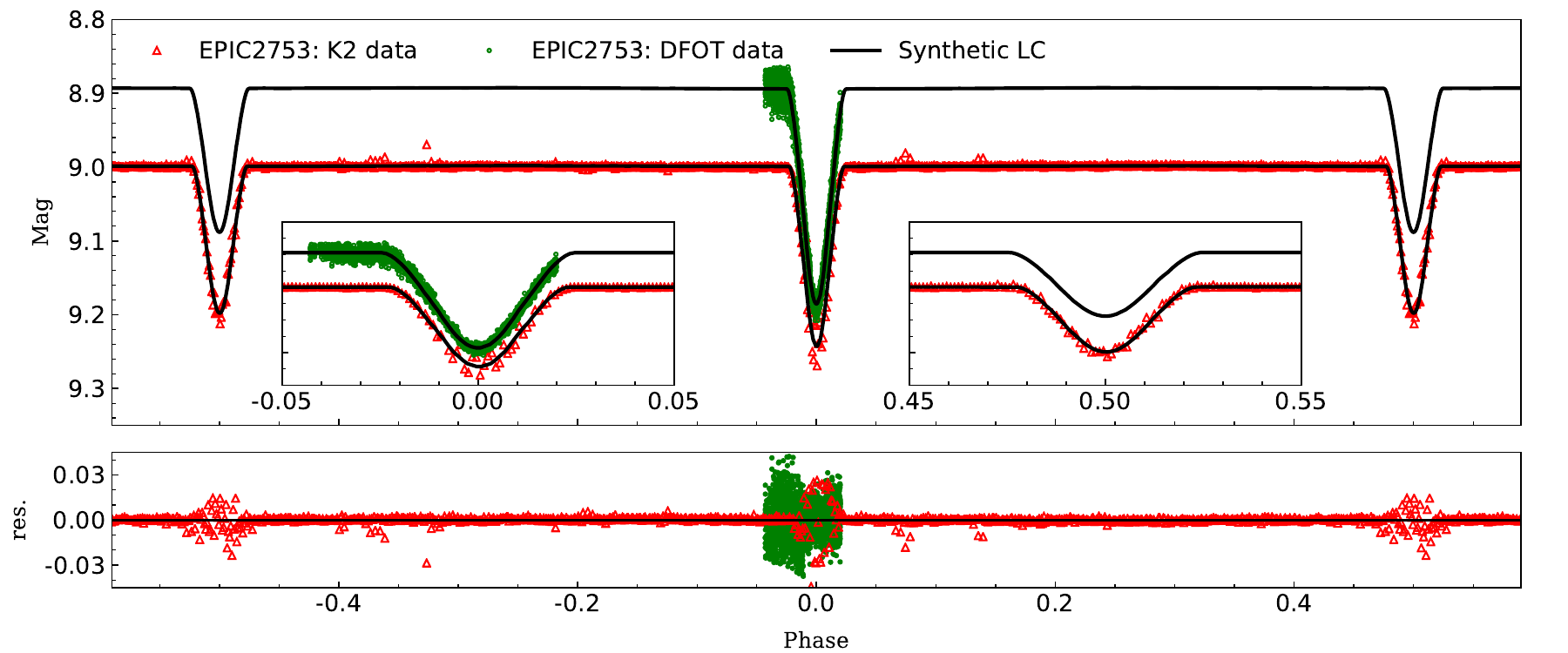}
\caption{The K2 mission observations (red triangles) and DFOT $R_{c}$ band observations (green dots) are shown with the synthetic light curve (black continuous line). The residuals between observed and synthetic data points are shown in the lower panel.}
\label{fig_03}
\end{center}
\end{figure}
\begin{figure}
\begin{center}
\includegraphics[width=7.5cm,height=7cm]{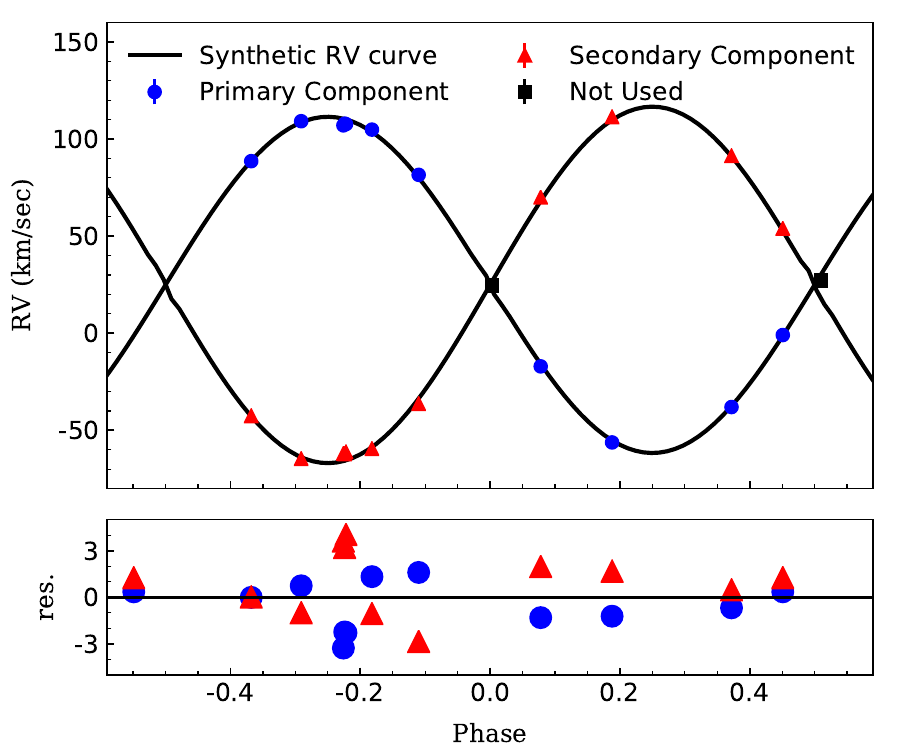}
\includegraphics[width=7.5cm,height=7cm]{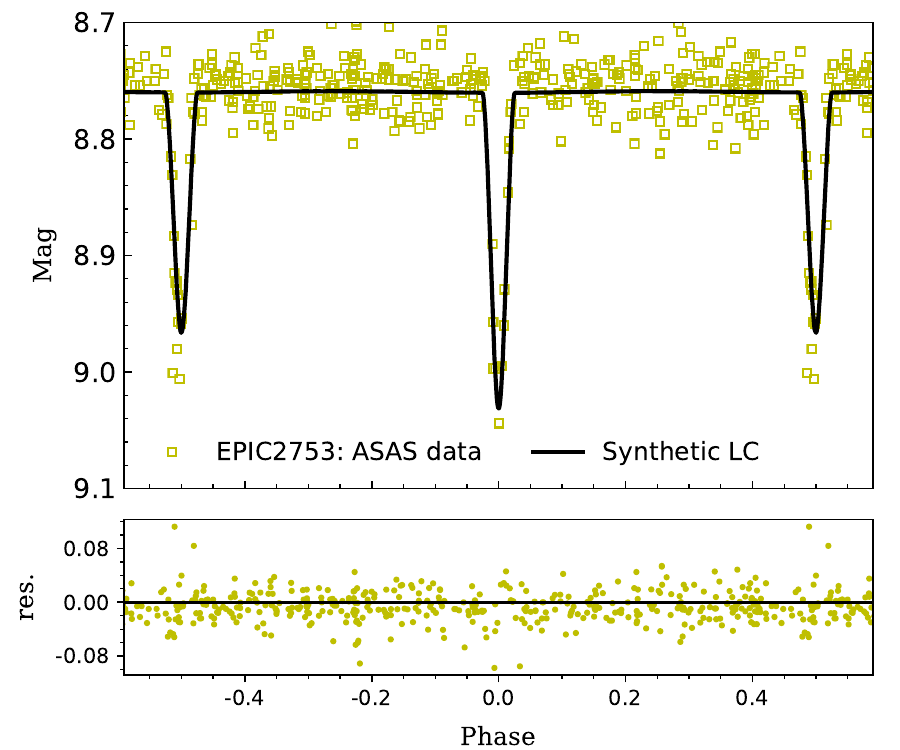}
\caption{RV variation for primary and secondary components along with synthetic RV curves are shown in the left panel. The ASAS-SN observations and the synthetic LC from PHOEBE are shown in the right panel.}
\label{fig_04}
\end{center}
\end{figure}

%
%
\textbf{Data Analysis: } To analyze the LCs and RV curves for the source, we used the LC and RV fitting package PHysics Of Eclipsing BinariEs (PHOEBE 1.0; \citealt{2005ApJ...628..426P}). Both the photometric data and RV curves can be analyzed at the same time in PHOEBE using its graphical user interface (GUI) or the PHOEBE-scripter. After loading the data in PHOEBE, we fixed the known parameters such as ToM at 0$^{th}$ orbital cycle, period of the system, and effective temperature ($T_{eff}$) of the primary component as determined by SED fitting. The eccentricity of the system was fixed to zero. Surface albedo and gravity brightening of components are fixed as 0.5 and 0.32. Limb darkening coefficients were updated using the tables from \cite{1993AJ....106.2096V} after each iteration. The fitted parameters were semi-major axis (a), mass-ratio (q), center of mass velocity ($V_{\gamma}$), inclination (i), secondary component temperature, primary star surface potential ($\Omega_{1}$), secondary star surface potential ($\Omega_{2}$), luminosity level ($l_{1}$). The detailed fitting procedure is described in \cite{2023MNRAS.521..677P}. The initial estimates of the best-fit parameters were obtained using the PHOEBE-GUI by analyzing the observed and synthetic data after every iteration. Parameters were further refined using the PHOEBE-scripter with multiple iterations. K2 and the DFOT LCs along with fitted LCs are shown in Figure~\ref{fig_03}. ASAS-SN LC is shown in the right panel of Figure~\ref{fig_04}. The left panel of Figure~\ref{fig_04} shows the RV curves for the primary and secondary components of the system. The data points close to the primary or secondary eclipse are not used in RV fitting due to single CCF peak (these include two observations from HERMES and one observation from APOGEE). The excluded points are represented by black squares in Figure~\ref{fig_04}.

The physical parameters are determined from the LC and RV modeling parameters using a FORTRAN-based code JKTABSDIM by \cite{2005A&A...429..645S}. The code requires information from LC and RV solutions such as period, RV amplitude, inclination, component temperatures, reddening, available photometry in different bands, etc. Based on the given information, this code determines the radius, mass, and luminosity of each component. For luminosity measurements, the code uses the bolometric correction information from \cite{2002A&A...391..195G}. The orbital solutions from PHOEBE and physical parameters for the system are given in Table~\ref{tab_01}. 

\begin{figure}
\begin{center}
\includegraphics[width=15cm,height=8cm]{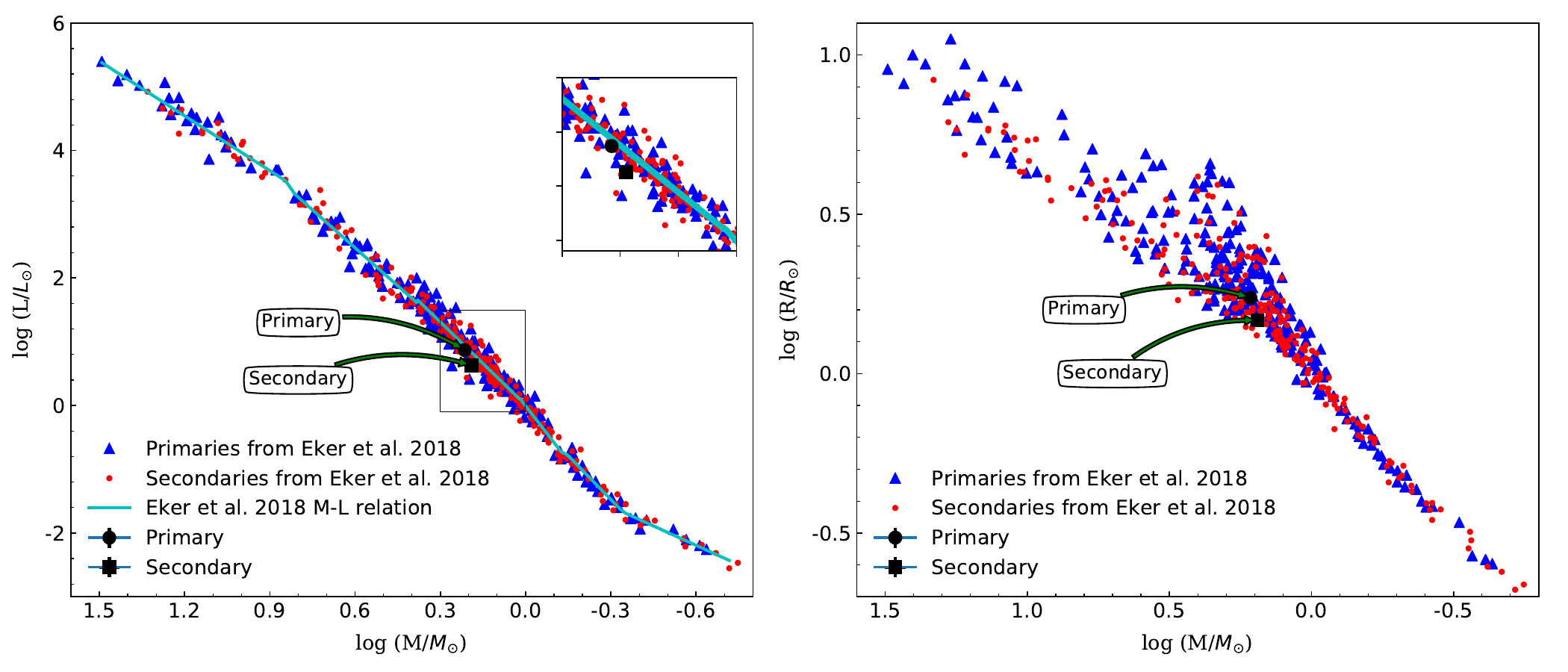}
\caption{EPIC 211982753 components along with other EBs from \cite{2018MNRAS.479.5491E}.}
\label{fig_05}
\end{center}
\end{figure}

\cite{2018MNRAS.479.5491E} derived parameter relations for EBs using a sample of 509 detached EBs. \cite{2018MNRAS.479.5491E} derived mass-luminosity and mass-radius relations using this sample. For our system the component luminosities are derived as $\log\,(L_{1})$=0.94(0.04) and $\log\,(L_{2})$=0.83(0.03) L$_{\odot}$ using the intermediate-mass stars mass-luminosity relation. The luminosity estimates by this relation are large as compared to our results. The EB sample used by \cite{2018MNRAS.479.5491E} is shown in Figure~\ref{fig_05}. 
The discrepancy in these estimates can be due to the non-homogeneous nature of the EB sample used by \cite{2018MNRAS.479.5491E}. A major fraction of the stars in the sample consists of main sequence stars from the solar neighborhood disc. Though actual metallicity estimates are unavailable for these systems, some works like \cite{2001MNRAS.325.1365H} and \cite{2021MNRAS.505.3165R} revealed the high metallicity among solar neighborhood stars. \cite{2018MNRAS.479.5491E} relations do not consider the effects of metallicity or age on output parameters. The behavior of mass-radius and mass-luminosity can change for stars with ages and metallicity.

\begin{table}       
\centering
\begin{tabular}{| l l || l l || l l |}    
\hline    
Parameters          & Results               & Parameters   & Results        & Parameters             & Results     \\
\hline \hline 
 & & & & & \\
a ($R_{\odot}$)     & 19.02(0.05)           & $\Omega_{1}$ & 13.56(0.11)    & $M_{1}$ ($M_{\odot}$)  & 1.64(0.02)  \\
$V_{\gamma}$ (km/sec)& 24.88(0.05)          & $\Omega_{2}$ & 12.88(0.15)    & $M_{2}$ ($M_{\odot}$)  & 1.55(0.01)  \\
q                   & 0.943(0.005)          & $r_{1}$ (a)  & 0.0907(0.0009) & $R_{1}$ ($R_{\odot}$)  & 1.73(0.02)  \\
i ($^{\circ}$)      & 85.14(0.08)           & $r_{2}$ (a)  & 0.0771(0.0010) & $R_{2}$ ($R_{\odot}$)  & 1.47(0.02)  \\
$T_{eff}^{pri}$	    & 7243 $(^{103}_{160}$) & $l_{1}$      & 8.08(0.09)     & log($L_{1}$)           & 0.87(0.03)  \\
$T_{eff}^{sec}$     & 6863(43)              & $l_{2}$      & 3.98           & log($L_{2}$)           & 0.63(0.02)  \\
 & & & & & \\
\hline                  
\end{tabular}
\caption{The combined LC and RV solutions for the system.}       
\label{tab_01}     
\end{table}

\section{Results and Discussion}
A new bright EB system  HD 69735 (or EPIC 211982753) was reported by \cite{2011MNRAS.416.2477W} using NASA's STEREO mission observations. The system was further observed by NASA's K2 mission during C-5 and C-18. We collected the available photometric data for the system using DFOT (only primary eclipse), ASAS-SN, and the K2 observations. We calculated the ToMs for the K2 time series and updated the ephemeris of the system. The O-C diagram is seen to follow a linear trend which is an indication of the non-variable period of the system. However, more photometric observations will further help to probe any long-term period changes in the system as the present analysis is based on a time length of $\sim$ three years only.

The available estimates of the effective temperature for the system in literature vary from 7000 (APOGEE survey) to 7500 K (GAIA survey DR3). The change in the effective temperature of any of the components can alter the estimated parameters like luminosity and radii. We used the SED fitting approach to determine the effective temperature of the components. The primary temperature from the SED fit was used to generate a template synthetic spectra for RV determination and data modeling. CCFs for each observed spectra were generated using the FXCOR routine of IRAF. RVs were determined by Gaussian fitting to the peaks in CCFs.

The photometric LCs and RV curves were solved simultaneously using the PHOEBE software. The parameters obtained from model fitting were further used to determine the fundamental parameters of the EB components using a FORTRAN-based code JKTABSDIM. The radii for primary and secondary components are determined as 1.73(0.02) and 1.47(0.02) $R_{\odot}$, respectively. The masses for primary and secondary components are determined as 1.64(0.02) and 1.55(0.01) $M_{\odot}$, respectively. The separation between the components is found to be 19.02(0.05) $R_{\odot}$. The distance to the system is determined as 238(4) pc while the distance estimate by GAIA-DR3 is 248(1) pc. The EB system is a total eclipsing system with an inclination angle of 85.14$^{\circ}$. The system is a high mass-ratio system with q=0.94.

We checked the position of EPIC 211982753 components on $M-T$, $M-L$, and $M-R\,$ planes using the MESA evolutionary tracks and isochrones. The primary component of the system was found to be within 160-225 Myr isochrones while the luminosity and radius of the secondary was lower than expected for this age range isochrones. Due to the unavailability of accurate metallicity estimates, we used isochrones with [Fe/H]=+0.02. The metallicity is mentioned in GAIA, LAMOST, and APOGEE catalogs but varies from one catalog to another. Accurate $\log g$ and metallicity estimates can result in a better age determination for the system in the future.

\begin{acknowledgments}
This work is supported by the Belgo-Indian Network for Astronomy and Astrophysics (BINA), approved by the International Division, Department of Science and Technology (DST, Govt. of India; DST/INT/BELG/P-09/2017) and the Belgian Federal Science Policy Office (BELSPO, Govt. of Belgium; BL/33/IN12). Guoshoujing Telescope (LAMOST) is a National Major Scientific Project built by the Chinese Academy of Sciences. Funding for the project has been provided by the National Development and Reform Commission. LAMOST is operated and managed by the National Astronomical Observatories, Chinese Academy of Sciences. In this paper, we have also used the data from the European Space Agency (ESA) mission Gaia, processed by the Gaia Data Processing and Analysis Consortium (DPAC). This work also makes use of the Two Micron All Sky Survey and SIMBAD database. This article is based on observations made in the Observatorios de Canarias del IAC with the Mercator telescope operated by the Institute of Astronomy, University of Leuven (Belgium) in the Observatorio del Roque de los Muchachos. The author acknowledges the accommodation and local travel support provided under the Belgo-Indian Network for Astronomy and Astrophysics (BINA) project for attending the BINA-3 workshop.
\end{acknowledgments}

\begin{furtherinformation}




\begin{authorcontributions}
Data acquisition, data analysis, and paper writing are done by the lead author of this paper while project and research methodology have been envisaged by Yogesh C. Joshi.
\end{authorcontributions}

\begin{conflictsofinterest}
The authors declare no conflict of interest.
\end{conflictsofinterest}

\end{furtherinformation}

\bibliographystyle{bullsrsl-en}

\bibliography{reference}

\end{document}